\documentclass[prd,twocolumn,10pt,a4paper,showpacs,preprintnumbers,amsmath,amssymb,notitlepage]{revtex4-1}
\date{March 2010}
\usepackage{epsfig}
\usepackage{latexsym}
\usepackage{amssymb}
\usepackage{booktabs}
\usepackage{graphicx}% Include figure files
\usepackage[utf8]{inputenc}
\newcommand{\be}{\begin{equation}}
\newcommand{\ee}{\end{equation}}
\newcommand{\ba}{\begin{eqnarray}}
\newcommand{\ea}{\end{eqnarray}}
\newcommand{\bi}{\begin{itemize}}
\newcommand{\ei}{\end{itemize}}

\newcommand{\<}{\langle}
\renewcommand{\>}{\rangle}
\newcommand{\eq}{Eq.~}

\newcommand{\fig}{Fig.~}

\newcommand{\la}{\label}

\newcommand{\txts}{\textstyle}

\begin{document}
\preprint{MITP/15-045}
\title{Lattice QCD calculation of hadronic light-by-light scattering}

\author{Jeremy Green$^a$, Oleksii Gryniuk$^{a,c}$, 
Georg von Hippel$^a$, Harvey~B.~Meyer$^{a,b}$, Vladimir Pascalutsa$^a$\medskip}

\affiliation{$^a$PRISMA Cluster of Excellence \& Institut f\"ur Kernphysik,
 \mbox{Johannes Gutenberg-Universit\"at Mainz,~D-55099~Mainz,~Germany}\\
 $^b$Helmholtz~Institut~Mainz, D-55099 Mainz, Germany\\
 $^c$Taras Shevchenko Kyiv National University, Volodymyrska 60, UA-01033 Kyiv, Ukraine\medskip}
 \email{\{green, hippel, meyerh, vladipas\}@kph.uni-mainz.de}
\medskip

\date{\today}

\begin{abstract}
\noindent We perform a lattice QCD calculation of the hadronic light-by-light scattering amplitude 
in a broad kinematical range.
At forward kinematics, the results are compared to a phenomenological analysis based on
dispersive sum rules for light-by-light scattering.
The  size of the pion pole contribution is investigated for momenta of typical hadronic size.
The presented numerical methods can be used to compute the hadronic light-by-light contribution 
to the anomalous magnetic moment of the muon.
Our calculations are carried out in two-flavor QCD with the pion mass in the range of 270 to 450$\,$MeV,
and contain so far only the diagrams with fully connected quark lines.
\end{abstract}

% \pacs{11.15.Ha, 11.25.Pm, 12.38.Aw} 
% { 12.38.Gc, 12.38.Mh, 25.75.-q}
\maketitle

\section{Introduction}

Light-by-light scattering, the elastic scattering of two photons, is a striking
prediction of Quantum Electrodynamics (QED). The light-by-light  (LbL) interaction appears prominently
in corrections to the anomalous magnetic moment ($g-2$) of the electron and muon. The muon
 $(g-2)$ exhibits a 3$\sigma$ discrepancy between experiment and the Standard
Model calculations~\cite{Blum:2013xva}. While the current theory and
experimental errors are comparable in size, a new $(g-2)_\mu$
experiment~\cite{Venanzoni:2014ixa} 
aiming to reduce the experimental error by a factor of four
is in preparation at Fermilab.  

The theory error on $(g-2)_\mu$ is dominated by hadronic
contributions, namely the hadronic vacuum polarization (HVP) and
hadronic light-by-light (HLbL) scattering.  Using unitarity and
causality, the HVP contribution is expressed in terms of the total
$e^+ e^- \to $ hadrons cross section, and hence its precision can
systematically be improved by collider experiments alone. By contrast,
the HLbL contribution cannot be expressed entirely in terms of cross
sections for $\gamma\gamma$-fusion into hadrons;
see~\cite{Pauk:2014rfa,Colangelo:2014dfa,Colangelo:2015ama} for
dispersive approaches to the problem.  A direct ab initio calculation
within Quantum Chromodynamics (QCD) is very challenging due to its
non-perturbative nature. In this work we address the problem using
lattice QCD.

A first lattice QCD+QED calculation of the HLbL contribution to
$(g-2)_\mu$ has recently been performed by Blum et
al.~\cite{Blum:2014oka}. We envisage a different method where the
four-point function for LbL scattering is computed in lattice QCD and
integrated over to yield the HLbL contribution.  In this Letter we
present the four-point function calculation and check it against the
available phenomenology. Exploiting unitarity and causality, the
forward HLbL amplitude can be expressed as a dispersive integral over
the $\gamma^*\gamma^*\to {\rm hadrons}$ cross
section~\cite{Pascalutsa:2010sj,Pascalutsa:2012pr}. A parametrization
of the latter allows us to confront the lattice calculation with
phenomenology in a fairly straightforward manner. 
As the neutral pion ($\pi^0$) pole dominates the HLbL
contribution to $(g-2)_\mu$ 
in phenomenological calculations~\cite{Blum:2013xva}, we study its
relative size both at forward and off-forward kinematics.

\vspace{-0.25cm}

\section{Theory background}

The Lehmann-Symanzik-Zimmermann reduction formula for the HLbL scattering amplitude implies~\footnote{
We use the notation and conventions of~\cite{Peskin:1995ev} unless 
otherwise stated. The metric is mostly minus. The fine-structure constant reads
$\alpha\equiv {e^2}/({4\pi})\simeq 1/137$.
The optical theorem for the scattering 
of scalar particles reads
${\rm Im}\,{\cal M}(p_1,p_2\to p_1,p_2)=2E_{\rm cm}p_{\rm cm}\sigma_{\rm tot}(p_1,p_2\to{\rm anything})$,
with $E_{\rm cm}$ the total center-of-mass energy and $p_{\rm cm}$ the norm of the three-momentum of one of the 
particles in the center-of-mass frame.}
\ba\la{eq:LSZ}
&&  {\cal M}_{\mu_1\mu_4\mu_2\mu_3}(p_1,p_4\to p_2,p_3) 
\\ && = e^4\,(-i\Pi_{\mu_1\mu_2\mu_3\mu_4}(-p_4;-p_1,p_2) ),
\nonumber
\ea
where $p_3 = p_1+p_4-p_2$ and 
\ba\la{eq:Pi4}
&& \Pi_{\mu_1\mu_2\mu_3\mu_4}(p_4;p_1,p_2) \equiv 
\int d^4x_1\,d^4x_2\,d^4x_4\,
\\ && e^{+i \sum_{a} p_a\cdot x_a}
 \big\<0\big|{\rm T}\big\{j_{\mu_1}(x_1)j_{\mu_2}(x_2)j_{\mu_3}(0)j_{\mu_4}(x_4)\big\}\big|0\big\>
\nonumber\ea
is the Minkowski-space time-ordered  correlator of the conserved vector current 
$j_\mu = \frac{2}{3} \bar u \gamma_\mu u - \frac{1}{3} \bar d \gamma_\mu d + \dots$.
The index $a$ takes the values 1, 2 and 4.
The components of the current $J_\mu$ used in the Euclidean theory~\footnote{
We use capital letters to denote `Euclidean' vectors, i.e.\ the metric in the scalar product
of two such vectors is understood to be Euclidean.}
are related  to their Minkowskian counterparts by $J_0 = j_0$,  $J_k = i\, j_k$.
The analytic continuation then yields the following relation to the Euclidean correlation function,
\ba\la{eq:Pi4a}
& & -i \Pi_{\mu_1\mu_2\mu_3\mu_4}\Big((-i{P_4^0},\vec P_4);(-i{P_1^0},\vec P_1),(-i{P_2^0},\vec P_2)\Big) 
\\ &&  = i^{n_0} \Pi^E_{\mu_1\mu_2\mu_3\mu_4}(P_4;P_1,P_2),\qquad
\nonumber
\\ 
&& \Pi^E_{\mu_1\mu_2\mu_3\mu_4}(P_4;P_1,P_2) \equiv 
\int d^4X_1 \;d^4X_2 \;d^4X_4 
\la{eq:Pi4b}\\ && e^{-i \sum_{a} P_a\cdot X_a}
\Big\<J_{\mu_1}(X_1)J_{\mu_2}(X_2)J_{\mu_3}(0)J_{\mu_4}(X_4)\Big\>_{\rm E},
\nonumber
\ea
where $n_0$ is the number of temporal indices carried by the vector currents 
in the correlator.

The forward scattering case is obtained in \eq(\ref{eq:LSZ}) by setting $p_2=p_1$.
Renaming the momenta to match the conventional notation, we have 
\ba\la{eq:Mforw}
&& {\cal M}_{\mu_1\mu_2\mu_3\mu_4}^{\rm forw}(q_1,q_2) \equiv
  {\cal M}_{\mu_1\mu_2\mu_3\mu_4}(q_1,q_2\to q_1,q_2) 
\\ && = e^4\,(-i\Pi_{\mu_1\mu_3\mu_4\mu_2}(-q_2;-q_1,q_1) ).
\nonumber
\ea
The forward scattering amplitude can be decomposed into eight Lorentz-invariant amplitudes~\cite{Budnev:1971sz}.
They are functions of the 
virtualities $q_1^2$ and $q_2^2$ of the photons, as well as of the variable $\nu\equiv q_1\cdot q_2$.
Using the projector $R^{\mu\nu}$ onto the subspace orthogonal to $q_1$ and $q_2$, 
we focus here on the amplitude~\footnote{In the notation of~\cite{Pascalutsa:2012pr}, 
${\cal M}_{\rm TT}  = \frac{1}{2}({M}_{++,++}+{M}_{+-,+-} )$ in terms of the helicity amplitudes.
By virtue of the optical theorem, 
the imaginary part of ${\cal M}_{\rm TT}$ is proportional to the total unpolarized $\gamma^*\gamma^*\to{\rm hadrons}$
cross-section. For the explicit expression of $R^{\mu\nu}$, see~\cite{Budnev:1971sz}.}
\be\la{eq:projM}
  {\cal M}_{\rm TT}(q_1^2,q_2^2,\nu) = \frac{1}{4} R^{\mu_1\mu_3}R^{\mu_2\mu_4}
{\cal M}_{\mu_1\mu_2\mu_3\mu_4}^{\rm forw}(q_1,q_2).
\ee
Combining Eqs.\ (\ref{eq:Mforw}) and (\ref{eq:Pi4a}), we can access the amplitude ${\cal M}_{\rm TT}$ 
from the Euclidean correlator,
\ba\la{eq:masterl}
&& {\cal M}_{\rm TT}(-Q_1^{\,2},-Q_2^{\,2},-Q_1\cdot Q_2)
\\ &=& \frac{ e^4}{4} R^E_{\mu_1\mu_3}R^E_{\mu_2\mu_4} \Pi^E_{\mu_1\mu_3\mu_4\mu_2}(-Q_2;-Q_1,Q_1),
\nonumber
\\
 R^E_{\mu\nu} &\equiv & \delta_{\mu\nu}  - \frac{1}{(Q_1\cdot Q_2)^2-Q_1^{\,2} Q_2^{\,2}} \cdot
\\ && \Big[(Q_1\cdot Q_2)(Q_1{}_\mu Q_2{}_\nu + Q_1{}_\nu Q_2{}_\mu)
\nonumber
\\ && 
 - Q_1^{\,2} Q_2{}_\mu Q_2{}_\nu - Q_2^{\,2} Q_1{}_\mu Q_1{}_\nu\Big].
\nonumber
\ea
The largest value of $|\nu|$ that can be reached with Euclidean kinematics is
limited by the virtualities of the photons~\footnote{One might be able to extend
the reach to $|\nu| = \nu_\pi$ with methods in the spirit of \cite{Ji:2001wha}.},
$ |\nu|  \leq  (Q_1^2 Q_2^2)^{1/2}\leq \frac{1}{2}(Q_1^2 + Q_2^2 )\equiv \nu_0$, while 
the nearest singularity is the s-channel $\pi^0$ pole located at
$ \nu_\pi = \frac{1}{2}(m_\pi^2 + Q_1^2 + Q_2^2)$. A technical issue arises 
when $Q_1$ and $Q_2$ are collinear: the projector $R^E_{\mu\nu}$ becomes ambiguous.
To resolve the issue, we note that $R^E_{\mu\nu} = \overline{R}_{\mu\nu} - U_1{}_\mu U_1{}_\nu$,
where $\overline{R}_{\mu\nu} \equiv \delta_{\mu\nu} - {Q_1{}_\mu Q_1{}_\nu}/Q_1^2$
and $U_1$ is the unit vector parallel to the projection of $Q_2$ onto the subspace orthogonal to $Q_1$.
The average of the applied projector over the directions of $U_1$ in that subspace yields
\ba
&&  \<\< R^E_{\mu_1\mu_3}R^E_{\mu_2\mu_4}\>\>_{U_1} 
= {\txts\frac{2}{5}} \overline{R}_{\mu_1\mu_3}\overline{R}_{\mu_2\mu_4}
\\ && \qquad \qquad 
+ {\txts\frac{1}{15}} \Big(\overline{R}_{\mu_1\mu_2} \overline{R}_{\mu_3\mu_4} 
+ \overline{R}_{\mu_1\mu_4} \overline{R}_{\mu_3 \mu_2} \Big).
\nonumber
\ea
We use this averaged projector in Eq.\ (\ref{eq:masterl}) when $Q_1$ and $Q_2$ are collinear.

In~\cite{Pascalutsa:2012pr}, it was shown that the HLbL amplitude ${\cal M}_{\rm TT}(\nu)$, 
for fixed spacelike photon virtualities, can be obtained from the following dispersive sum rule,
\ba\la{eq:dr}
&& {\cal M}_{\rm TT}(q_1^2,q_2^2,\nu) - {\cal M}_{\rm TT}(q_1^2,q_2^2,0) 
\\ &&  = \frac{2\nu^2}{\pi}
\int_{\nu_0}^\infty d\nu'\frac{ \sqrt{ \nu'{}^2 - q_1^2 q_2^2  }}{\nu'(\nu'{}^2-\nu^2-i\epsilon)}(\sigma_0+\sigma_2)(\nu'),
\nonumber
\ea
where $\sigma_0$ and $\sigma_2$ are the total cross sections $\gamma^*(q_1^{\,2})\gamma^*(q_2^{\,2})\to {\rm hadrons}$ with 
total helicity 0 and 2 respectively. It can be shown~\cite{Pascalutsa:2012pr} that
${\cal M}_{\rm TT}$ vanishes at $\nu=0$ if either of the photons is real.
It is interesting to test the sum rule for the $\pi^0$ pole contribution.
Using the expression for $\Pi_{\mu\nu\rho\sigma}$ given in~\cite{Knecht:2001qf} and Eqs.\ (\ref{eq:Mforw}, \ref{eq:projM}),
one finds
\ba\la{eq:Mpi0tot}
&& {\cal M}_{\rm TT}^{\pi^0}(-Q_1^2,-Q_2^2,\nu) = e^4 \,(\nu{}^2 - Q_1^2 Q_2^2  )\,
\\ 
&& \qquad {\cal F}(-Q_1^2,-Q_2^2)^2  \frac{Q_1^2+Q_2^2+m_\pi^2}{(Q_1^2+Q_2^2+m_\pi^2)^2-4\nu^2}
\nonumber
\ea
with ${\cal F}(q_1^2,q_2^2)$ the pion transition form factor as defined in~\cite{Knecht:2001qf}.
For $q_2^2=0$, the same result is obtained from the sum rule, using the expression for the 
$\gamma\gamma^*\to \pi^0$ cross-section given in~\cite{Pascalutsa:2012pr}.

In summary, the amplitude ${\cal M}_{\rm TT}$ can be computed on the lattice 
via \eq (\ref{eq:masterl}) and from $e^+ e^-$ collider data via \eq (\ref{eq:dr}).
In the following, we present a comparison of the two approaches.

\vspace{-0.3cm}

\section{Implementation of the Euclidean four-point function in lattice QCD}

\begin{figure}
  \centering
  \includegraphics[width=\columnwidth]{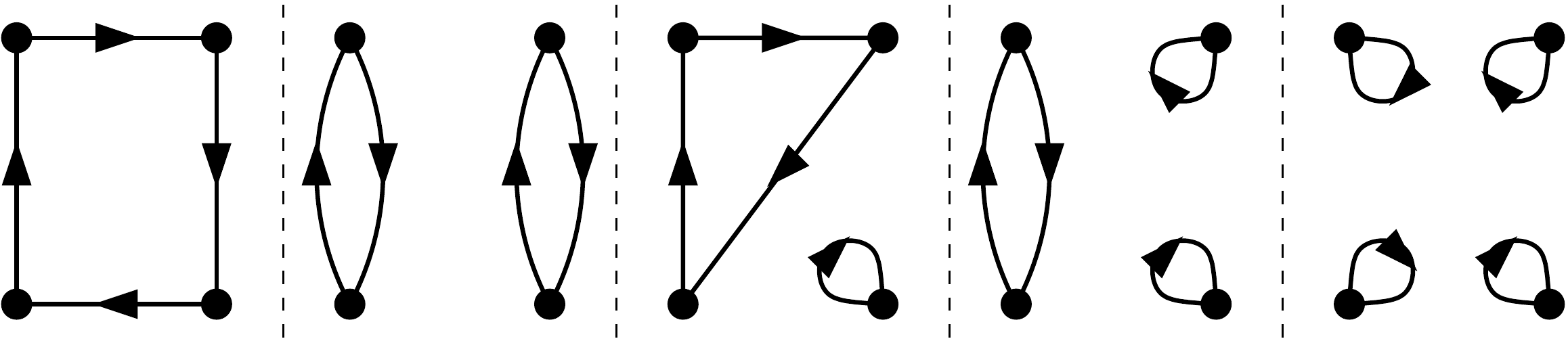}
  \caption{Four-point function quark contraction topologies. The
    vertices represent vector currents and the lines are quark
    propagators. In this work, we compute only the leftmost,
    fully-connected class of diagrams.}
  \label{fig:contractions}
\end{figure}

In numerical lattice QCD calculations of $n$-point functions, the
quark path integral is evaluated analytically to yield a sum of
contractions of quark propagators. For the four-point function of
vector currents, these fall into five distinct topologies, illustrated
in Fig.~\ref{fig:contractions}. In this work, we compute only the six
contractions that are fully quark-connected.

\begin{figure}
  \centering
  \includegraphics[width=0.75\columnwidth]{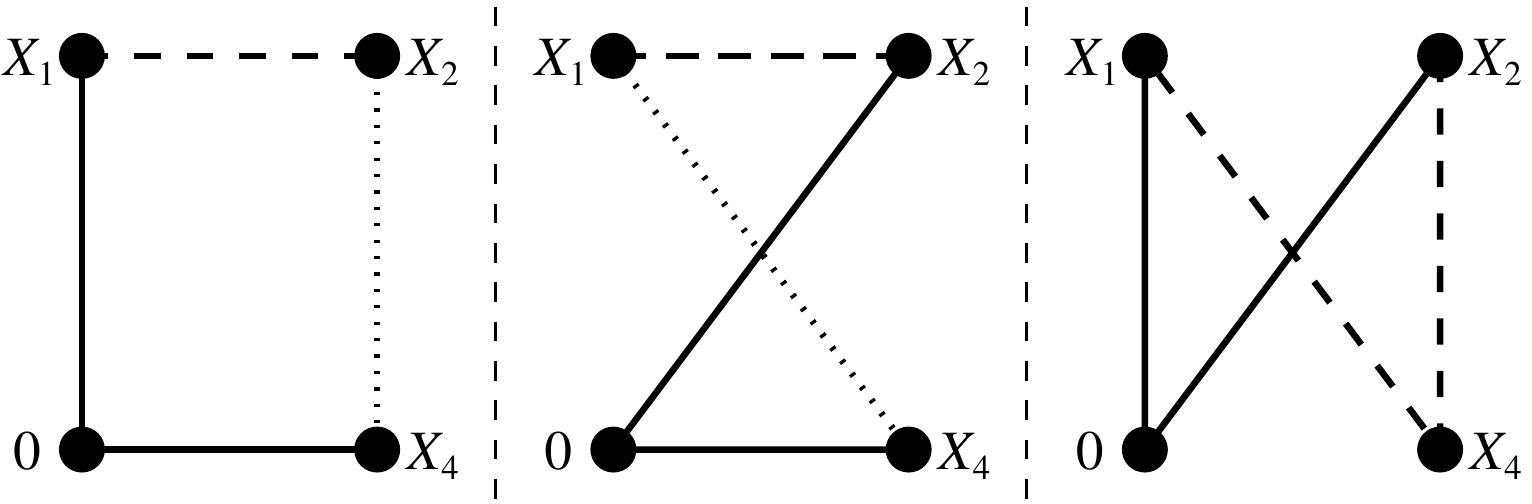}
  \caption{Fully-connected four-point function quark
    contractions. Each panel represents two contractions with opposite
    directions of quark flow. The solid quark lines are computed using
    a point-source propagator, the dashed lines using sequential
    propagators, and the dotted lines using double-sequential
    propagators.}
  \label{fig:connected}
\end{figure}

We use a Wilson-type quark action, 
three lattice conserved currents $J_\mu^c$ and one site-local
current $J_\mu^l$ (see for instance~\cite{Francis:2013fzp} for an explicit definition).
Generically, we evaluate the fully-connected contribution to
\begin{multline}
  \label{eq:pilat}
  \Pi^\text{lat}_{\mu_1\mu_2\mu_3\mu_4}(X_4;f_1,f_2)\equiv
\sum_{X_1,X_2}f_1(X_1)f_2(X_2)\\
\< J_{\mu_1}^c(X_1) J_{\mu_2}^c(X_2) J_{\mu_3}^l(0) J_{\mu_4}^c(X_4) + \text{contact terms} \>,
\end{multline}
for some fixed functions $f_{1,2}$ and all values of $\{\mu_a\}$ and
$ X_4$. The contact terms are present when two or three lattice
conserved currents coincide, and serve to ensure that the
conserved-current relations hold, e.g.,
$\Delta^{(X_4)}_{\mu_4}\Pi^\text{lat}_{\mu_1\mu_2\mu_3\mu_4} = 0$,
where $\Delta_\mu^{(X)}$ is the backward lattice derivative.

The fully-connected contribution to Eq.~(\ref{eq:pilat}) is evaluated
using the method of sequential propagators. First, a point-source
propagator is computed from $X_3$. Then, it is combined with the
function $f_1$ or $f_2$ to form the source for a new (sequential)
propagator. These sequential propagators are then used to form sources
for double-sequential propagators that depend on both $f_1$ and
$f_2$. Finally, the fully-connected contraction is formed using all
three kinds of propagators; this is illustrated in
Fig.~\ref{fig:connected}. For generic complex $f_1$ and $f_2$, this
requires one point-source, 16 sequential and 32 double-sequential
propagators, although these counts can be reduced in various special
cases. We have verified that in our implementation  the four-point
function  matches the lattice perturbation theory
calculation if the gauge link variables are set to unity,
and that the conserved-current conditions hold on each gauge
configuration.

For evaluating the momentum-space correlator, we set the functions to
be plane waves, $f_a(X)=e^{-iP_a\cdot X}$ and compute the Fourier
modes with respect to $X_4$. Thus,
$\Pi^E_{\mu_1\mu_2\mu_3\mu_4}(P_4;P_1,P_2)$ can be 
evaluated efficiently at fixed $P_{1,2}$ for all $P_4$ available on the lattice.

\begin{figure}[t!]
\centerline{\includegraphics*[width=0.45\textwidth]{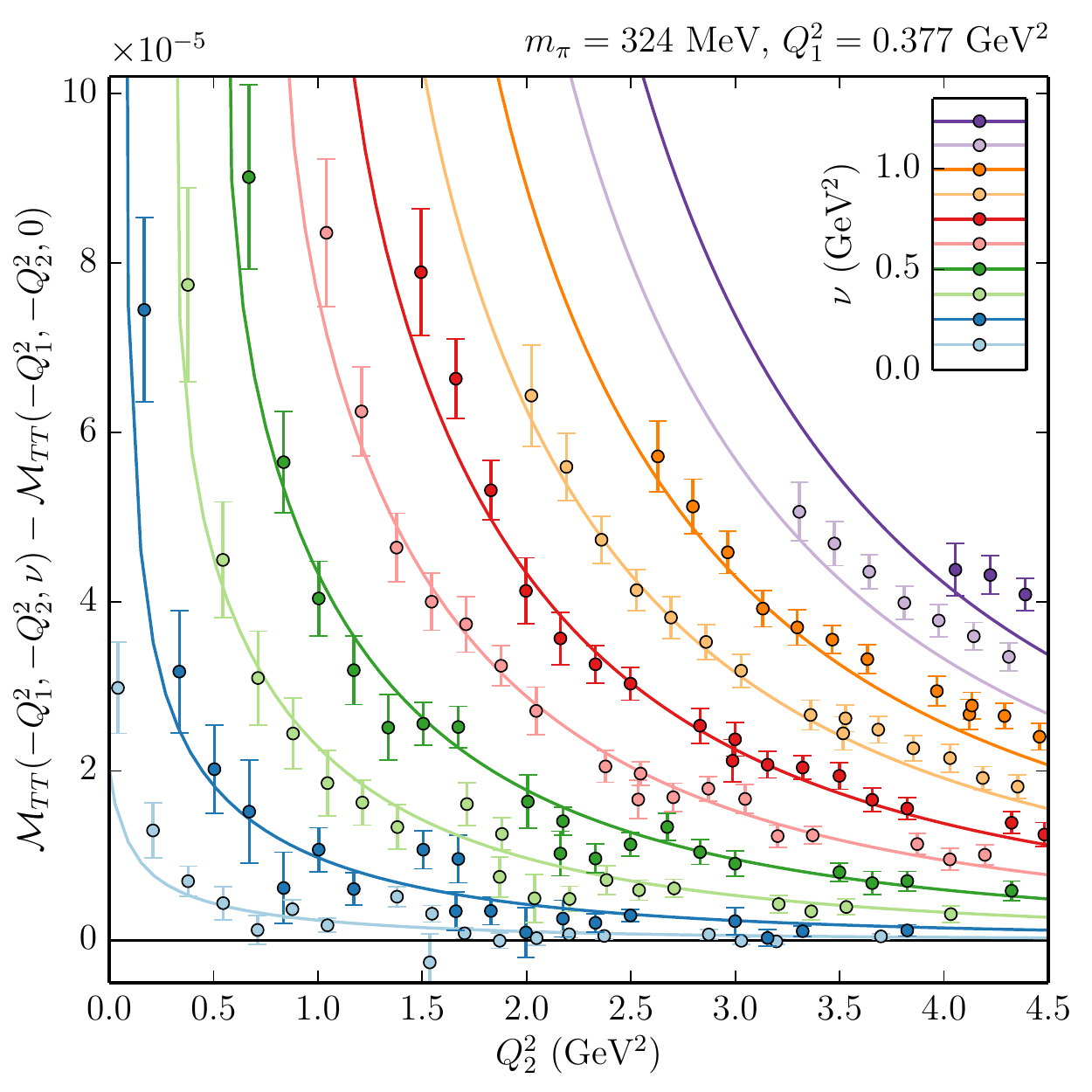}}
\vspace{-0.4cm}
\caption{\label{fig:Q2_dep} The forward scattering amplitude ${\cal M}_{\rm TT}$ 
at a fixed virtuality $Q_1^2=0.377{\rm GeV}^2$, as a function of the other photon virtuality $Q_2^2$,
for different values of $\nu$. The curves represent the predictions based on Eq.\ (\ref{eq:dr}), see the text for details.}
\end{figure}

\section{Results}

We have used three lattice QCD ensembles with two degenerate flavors
of non-perturbatively O($a$) improved Wilson quarks and a plaquette
gauge action.  The ensembles are at a single lattice spacing
$a=0.063{\rm fm}$~\cite{Capitani:2011fg}, correspond to pion masses
$m_\pi = 451,~324$ and 277\,MeV, and are respectively of
spatial linear size 32, 48 and 48,  the time direction being twice as
long; see~\cite{Fritzsch:2012wq} for more
details. Only the up and down quark contributions to the electromagnetic current
are included. The local vector current $J_\mu^l$ is renormalized
non-perturbatively~\cite{DellaMorte:2005rd}. The results shown here
were obtained using fairly low statistics, with a maximum of 300
samples.

\begin{figure}[t!]
\centerline{\includegraphics*[width=0.45\textwidth]{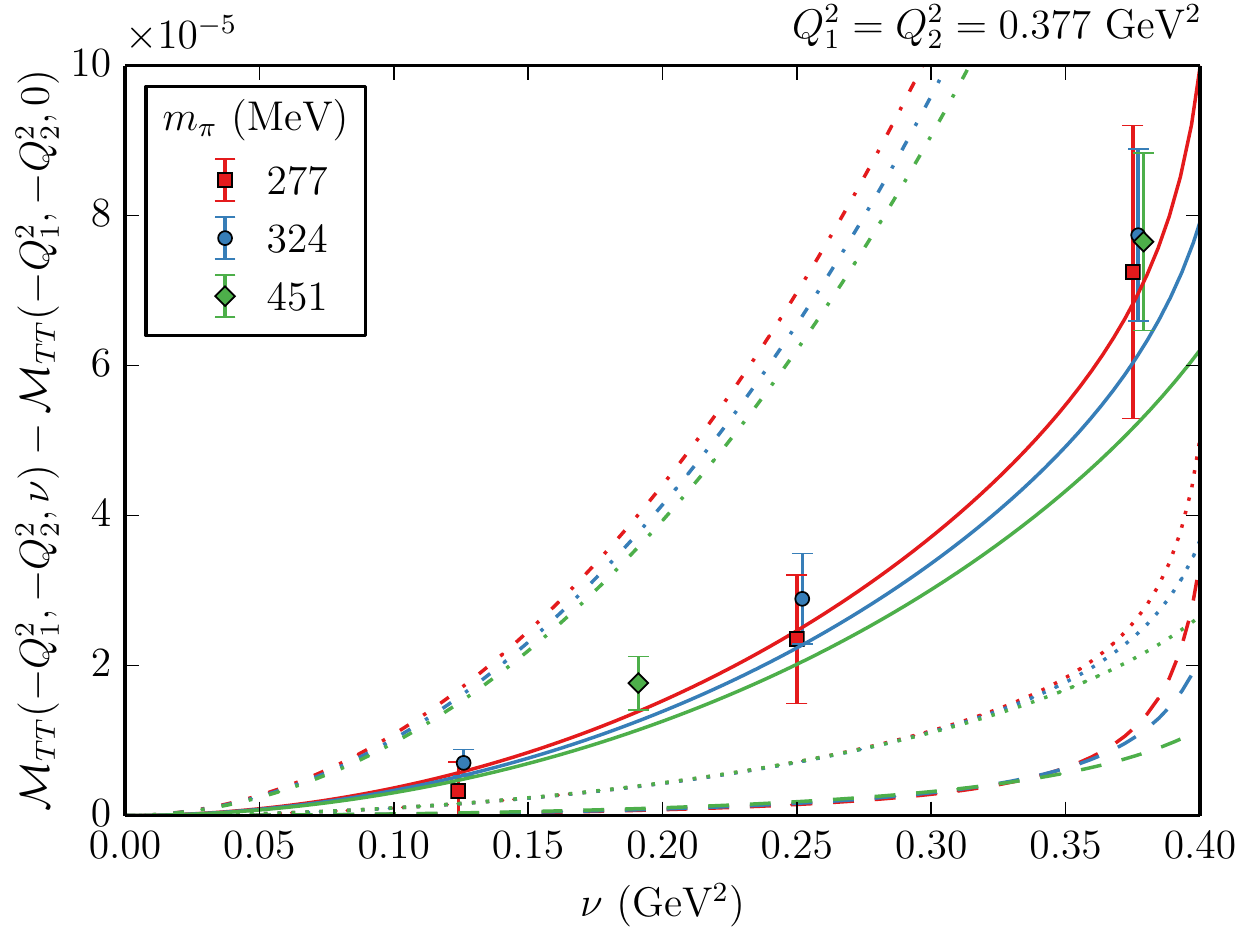}}
\vspace{-0.4cm}
\caption{\label{fig:mpi_dep} The dependence of the amplitude ${\cal
    M}_{\rm TT}$ on $\nu$, both photon virtualities being fixed at
  0.377~GeV$^2$, at three different pion masses. The dashed and dotted
  curves show the $\pi^0$ and $\pi^0+\eta'$ contributions 
  (there is no $\eta$ meson in two-flavor QCD), the solid
  curve includes all single-meson and $\pi^+\pi^-$ contributions, and
  the dash-dotted curves additionally include the high-energy
  contribution for the case of real photons at the physical pion
  mass.}
\end{figure}

Due to the finite volume of the lattice, the momenta take discrete values.
The subtracted forward scattering amplitude,
$\mathcal{M}_{\rm TT}(-Q_1^2,-Q_2^2,\nu)-\mathcal{M}_{\rm TT}(-Q_1^2,-Q_2^2,0)$
(which is even in $\nu$), is obtained by linearly interpolating the
second term between the available $Q_2^2$ to match the first term. It
is shown in Fig.~\ref{fig:Q2_dep} at fixed pion mass and fixed $Q_1^2$, and
also in Fig.~\ref{fig:mpi_dep} with both photon virtualities
fixed. For the latter, linear interpolation in $Q_2^2$ was also used
in the first term, except for the points at maximal $\nu$. At fixed
$\nu$, the amplitude tends to decrease as the virtualities are
increased, at fixed virtualities it tends to increase with $|\nu|$,
and at fixed kinematics we do not find a significant dependence on the
pion mass.

\begin{figure}[t!]
\centerline{\includegraphics*[width=0.45\textwidth]{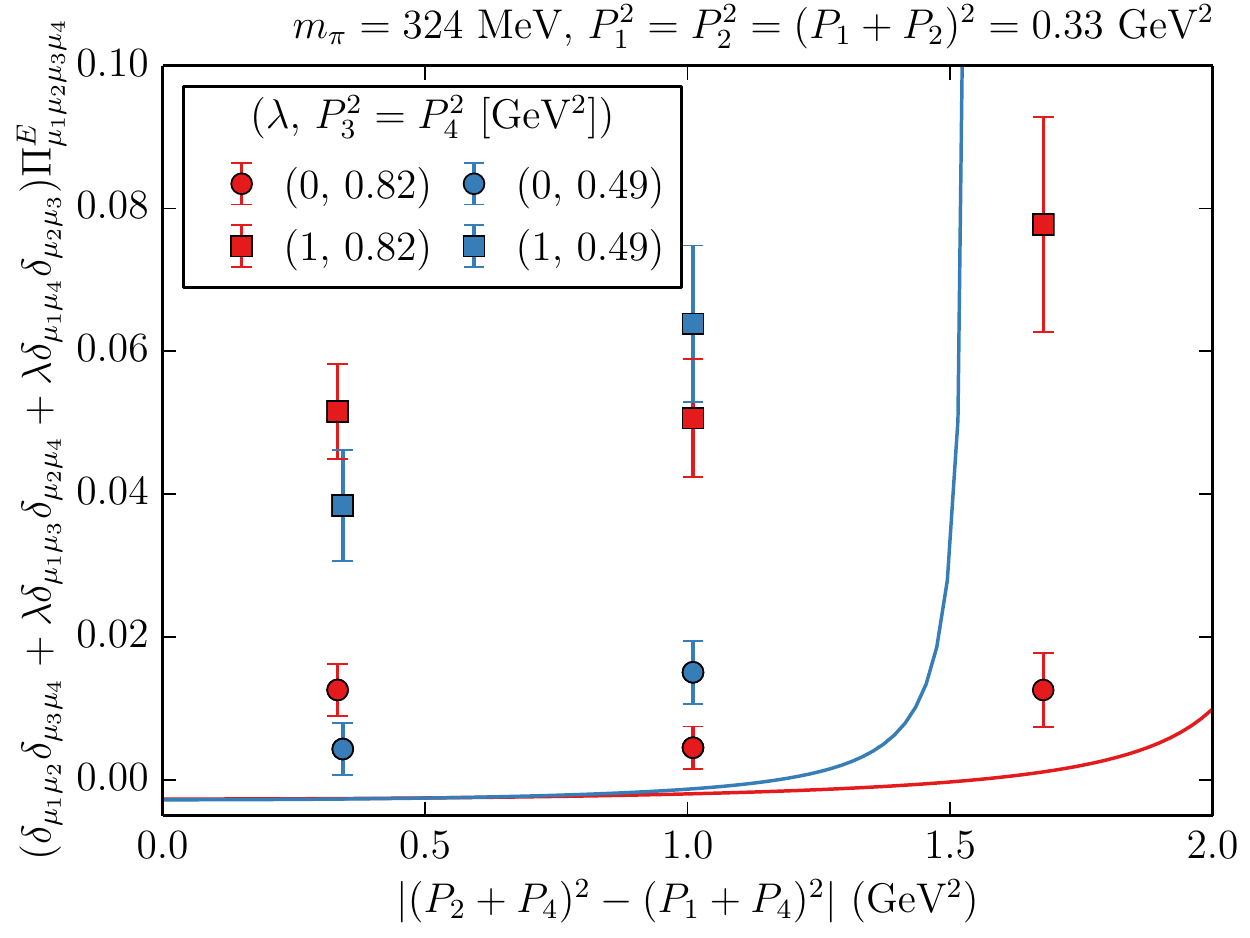}}
\vspace{-0.4cm}
\caption{\label{fig:genkin} Two Lorentz contractions of the vector four-point function at non-forward kinematics.
For $\lambda=1$ (squares), the pion pole contribution vanishes, while for $\lambda=0$ (circles), it  does not.
The curves correspond to the $\pi^0$ pole contribution in the latter case.}
\end{figure}

We compare the lattice data with results from the sum rule,
Eq.~(\ref{eq:dr}), using a phenomenological model for the transverse
$\gamma^*\gamma^*\to\text{hadrons}$ cross section,
$\sigma_0+\sigma_2$, based on Ref.~\cite{PhysRevD.90.036004}. We
include pseudoscalar, scalar, axial-vector, and tensor mesons, as well
as the non-resonant $\pi^+\pi^-$ contribution  % ~\cite{Pascalutsa:2012pr} 
(in scalar treelevel QED with pion electromagnetic form   % {PhysRevD.90.036004}
factors). The $\gamma^*\gamma^*\to\text{meson}$ form factors have not
been measured experimentally; they are assumed to factorize as
$F(q_1^2,q_2^2)=F(q_1^2,0)F(0,q_2^2)/F(0,0)$. For the pseudoscalar and
axial-vector mesons, $F(q^2,0)=F(0,q^2)$ is described based on
experimental data as in Ref.~\cite{Pascalutsa:2012pr} and, lacking
guidance from experiment, we assume a monopole form factor for the
scalar and tensor resonances with a pole mass set by hand to
$\Lambda=1.6$~GeV. The model is modified for unphysical quark masses
by adjusting the masses and $\gamma\gamma$ decay widths,
$\Gamma_{\gamma\gamma}$, of the mesons. The pion mass and decay
constant~\cite{Engel:2014cka} $f_\pi$ are calculated on each lattice
ensemble, and  ${\cal F}(0,0)$ is set to the value $(-4\pi^2 f_\pi)^{-1}$
inspired by the chiral anomaly
prediction (see e.g.\ \cite{Bernstein:2011bx}). 
For each of the remaining mesons, the mass is assumed to have the same
shift as that of the $\rho$ meson, relative to the physical point, and
$\Gamma_{\gamma\gamma}$ is assumed to scale linearly with the meson's
mass.

This model together with the dispersive sum rule produces the solid
curves in Figs.~\ref{fig:Q2_dep} and \ref{fig:mpi_dep}, which agree
well with the data. Varying $\Lambda$ by $\pm0.4$~GeV shifts the
curves by up to $\pm50\%$, hence it is clear that the model has a
considerable uncertainty; nevertheless the consistency with the
data is remarkable. Fig.~\ref{fig:mpi_dep} also shows the individual contributions
from $\pi^0$ and $\eta'$ mesons and a high-energy contribution arising from a fit to
the total $\gamma\gamma\to\text{hadrons}$ cross
section~\cite{Agashe:2014kda} based on Regge theory. The latter is
excluded from the main model curves due to the lack of a
well-motivated extra\-po\-lation to the case of virtual photons and
larger-than-physical pion masses.
It is interesting to note that the two-pion production is 
typically the dominant contribution to the amplitude, rather than the
$\pi^0$ and $\eta'$ production.

Moving to off-forward kinematics, the situation is more
complicated. In general, the four-point function of vector currents
can be decomposed into 41 Lorentz-invariant
functions~\cite{Eichmann:2015nra} (see also \cite{Colangelo:2015ama})
that depend on six kinematic variables, of which three are fixed when
$P_1$ and $P_2$ are fixed in our lattice calculation. To study the
importance of the $\pi^0$ contribution, we consider two contractions:
$\Pi^E_{\mu_1\mu_1\mu_3\mu_3}$, which has pion poles when
$(P_1+P_4)^2=-m_\pi^2$ or $(P_2+P_4)^2=-m_\pi^2$, and a
fully-symmetric contraction that has no $\pi^0$-exchange
contribution. These are shown in Fig.~\ref{fig:genkin}, where we have
also fixed $P_3^2=P_4^2$ [where $P_3=-(P_1+P_2+P_4)$] to be a typical
hadronic scale below 1~GeV$^2$. We find that the fully-symmetric
contraction yields larger data, again indicating that the $\pi^0$ does
not provide the dominant contribution.

\section{Conclusion}

We have demonstrated that the fully connected contribution to the
momentum-space four-point function of the electromagnetic current can
be computed with mo\-derate computational effort in lattice QCD if two
of the three momenta are fixed.  As an application, we computed one of
the forward $\gamma^*\gamma^*$ scattering amplitudes in a broad
kinematic range. Via a dispersive sum rule, it is related
model-independently to $\gamma^*\gamma^*\to{\rm hadrons}$ cross
sections.  Modelling the latter, we find the comparison of the lattice
calculation with the phenomenological approach to be successful.  
The systematic uncertainties of the comparison are presently still large,
mainly because our current calculations are performed at heavier quark
masses than the physical ones, but this model dependence can be
syste\-ma\-tically reduced. Also, the not fully connected contraction
topologies depicted in \fig\ref{fig:contractions} could be important.
We investigated the size of the pion pole contribution both in the
forward and the off-forward amplitude.  Both the lattice data and the
model show that it is by no means dominant in a range of kinematic
invariants of typical hadronic size.

The numerical methods presented can be applied to a direct lattice
calculation of the HLbL contribution to $(g-2)_\mu$: we are currently
working on a position-space approach where the photon propagators are
integrated out semi-analytically in infinite volume.  The dominant
systematic effects are likely to be quite different from those in the
method of Blum et al.~\cite{Blum:2014oka}, allowing for useful
cross-checks.  Since phenomenological calculations indicate that the
$\pi^0$ is dominant in the HLbL contribution to
$(g-2)_\mu$~\cite{Blum:2013xva}, realistically light quark masses and
large volumes will be required to treat this long-range contribution
correctly. Lattice data on the HLbL amplitude itself can also help
discriminate between pheno\-me\-no\-logical models used in the calculation
of $(g-2)_\mu$.

\begin{acknowledgments}
\noindent We thank A.\ Nyffeler, M.\ Vanderhaeghen  and H.~Wittig
for helpful discussions. We are grateful for the access to the
CLS lattice ensembles used here. 
The correlation functions were computed on the
`{Clover}' cluster at the Helmholtz-Institut Mainz.
The programs were written using QDP++~\cite{Edwards:2004sx} with
the deflated SAP+GCR solver from openQCD~\cite{OpenQCD}.
\end{acknowledgments}

%%%%%%%%%%%%%%%%%%%%%%%%%%%%%%%%%%%%%%%%%%%%%%%%%%%%%%%%%%%%%%%%
%\bibliography{/Users/harvey/BIBLIO/viscobib}
\bibliography{fhlbl}

\begin{thebibliography}{27}%
\makeatletter
\providecommand \@ifxundefined [1]{%
 \@ifx{#1\undefined}
}%
\providecommand \@ifnum [1]{%
 \ifnum #1\expandafter \@firstoftwo
 \else \expandafter \@secondoftwo
 \fi
}%
\providecommand \@ifx [1]{%
 \ifx #1\expandafter \@firstoftwo
 \else \expandafter \@secondoftwo
 \fi
}%
\providecommand \natexlab [1]{#1}%
\providecommand \enquote  [1]{``#1''}%
\providecommand \bibnamefont  [1]{#1}%
\providecommand \bibfnamefont [1]{#1}%
\providecommand \citenamefont [1]{#1}%
\providecommand \href@noop [0]{\@secondoftwo}%
\providecommand \href [0]{\begingroup \@sanitize@url \@href}%
\providecommand \@href[1]{\@@startlink{#1}\@@href}%
\providecommand \@@href[1]{\endgroup#1\@@endlink}%
\providecommand \@sanitize@url [0]{\catcode `\\12\catcode `\$12\catcode
  `\&12\catcode `\#12\catcode `\^12\catcode `\_12\catcode `\%12\relax}%
\providecommand \@@startlink[1]{}%
\providecommand \@@endlink[0]{}%
\providecommand \url  [0]{\begingroup\@sanitize@url \@url }%
\providecommand \@url [1]{\endgroup\@href {#1}{\urlprefix }}%
\providecommand \urlprefix  [0]{URL }%
\providecommand \Eprint [0]{\href }%
\providecommand \doibase [0]{http://dx.doi.org/}%
\providecommand \selectlanguage [0]{\@gobble}%
\providecommand \bibinfo  [0]{\@secondoftwo}%
\providecommand \bibfield  [0]{\@secondoftwo}%
\providecommand \translation [1]{[#1]}%
\providecommand \BibitemOpen [0]{}%
\providecommand \bibitemStop [0]{}%
\providecommand \bibitemNoStop [0]{.\EOS\space}%
\providecommand \EOS [0]{\spacefactor3000\relax}%
\providecommand \BibitemShut  [1]{\csname bibitem#1\endcsname}%
\let\auto@bib@innerbib\@empty
%</preamble>
\bibitem [{\citenamefont {Blum}\ \emph {et~al.}(2013)\citenamefont {Blum},
  \citenamefont {Denig}, \citenamefont {Logashenko}, \citenamefont {de~Rafael},
  \citenamefont {Lee~Roberts} \emph {et~al.}}]{Blum:2013xva}%
  \BibitemOpen
  \bibfield  {author} {\bibinfo {author} {\bibfnamefont {T.}~\bibnamefont
  {Blum}}, \bibinfo {author} {\bibfnamefont {A.}~\bibnamefont {Denig}},
  \bibinfo {author} {\bibfnamefont {I.}~\bibnamefont {Logashenko}}, \bibinfo
  {author} {\bibfnamefont {E.}~\bibnamefont {de~Rafael}}, \bibinfo {author}
  {\bibfnamefont {B.}~\bibnamefont {Lee~Roberts}},  \emph {et~al.},\
  }\href@noop {} {\  (\bibinfo {year} {2013})},\ \Eprint
  {http://arxiv.org/abs/1311.2198} {arXiv:1311.2198 [hep-ph]} \BibitemShut
  {NoStop}%
%%CITATION = ARXIV:1311.2198;%%
\bibitem [{\citenamefont {Venanzoni}(2014)}]{Venanzoni:2014ixa}%
  \BibitemOpen
  \bibfield  {author} {\bibinfo {author} {\bibfnamefont {G.}~\bibnamefont
  {Venanzoni}} (\bibinfo {collaboration} {Muon g-2}),\ }\href@noop {} {\
  (\bibinfo {year} {2014})},\ \Eprint {http://arxiv.org/abs/1411.2555}
  {arXiv:1411.2555 [physics.ins-det]} \BibitemShut {NoStop}%
%%CITATION = ARXIV:1411.2555;%%
\bibitem [{\citenamefont {Pauk}\ and\ \citenamefont
  {Vanderhaeghen}(2014)}]{Pauk:2014rfa}%
  \BibitemOpen
  \bibfield  {author} {\bibinfo {author} {\bibfnamefont {V.}~\bibnamefont
  {Pauk}}\ and\ \bibinfo {author} {\bibfnamefont {M.}~\bibnamefont
  {Vanderhaeghen}},\ }\href {\doibase 10.1103/PhysRevD.90.113012} {\bibfield
  {journal} {\bibinfo  {journal} {Phys.Rev.}\ }\textbf {\bibinfo {volume}
  {D90}},\ \bibinfo {pages} {113012} (\bibinfo {year} {2014})},\ \Eprint
  {http://arxiv.org/abs/1409.0819} {arXiv:1409.0819 [hep-ph]} \BibitemShut
  {NoStop}%
%%CITATION = ARXIV:1409.0819;%%
\bibitem [{\citenamefont {Colangelo}\ \emph {et~al.}(2014)\citenamefont
  {Colangelo}, \citenamefont {Hoferichter}, \citenamefont {Procura},\ and\
  \citenamefont {Stoffer}}]{Colangelo:2014dfa}%
  \BibitemOpen
  \bibfield  {author} {\bibinfo {author} {\bibfnamefont {G.}~\bibnamefont
  {Colangelo}}, \bibinfo {author} {\bibfnamefont {M.}~\bibnamefont
  {Hoferichter}}, \bibinfo {author} {\bibfnamefont {M.}~\bibnamefont
  {Procura}}, \ and\ \bibinfo {author} {\bibfnamefont {P.}~\bibnamefont
  {Stoffer}},\ }\href {\doibase 10.1007/JHEP09(2014)091} {\bibfield  {journal}
  {\bibinfo  {journal} {JHEP}\ }\textbf {\bibinfo {volume} {1409}},\ \bibinfo
  {pages} {091} (\bibinfo {year} {2014})},\ \Eprint
  {http://arxiv.org/abs/1402.7081} {arXiv:1402.7081 [hep-ph]} \BibitemShut
  {NoStop}%
%%CITATION = ARXIV:1402.7081;%%
\bibitem [{\citenamefont {Colangelo}\ \emph {et~al.}(2015)\citenamefont
  {Colangelo}, \citenamefont {Hoferichter}, \citenamefont {Procura},\ and\
  \citenamefont {Stoffer}}]{Colangelo:2015ama}%
  \BibitemOpen
  \bibfield  {author} {\bibinfo {author} {\bibfnamefont {G.}~\bibnamefont
  {Colangelo}}, \bibinfo {author} {\bibfnamefont {M.}~\bibnamefont
  {Hoferichter}}, \bibinfo {author} {\bibfnamefont {M.}~\bibnamefont
  {Procura}}, \ and\ \bibinfo {author} {\bibfnamefont {P.}~\bibnamefont
  {Stoffer}},\ }\href@noop {} {\  (\bibinfo {year} {2015})},\ \Eprint
  {http://arxiv.org/abs/1506.01386} {arXiv:1506.01386 [hep-ph]} \BibitemShut
  {NoStop}%
%%CITATION = ARXIV:1506.01386;%%
\bibitem [{\citenamefont {Blum}\ \emph {et~al.}(2015)\citenamefont {Blum},
  \citenamefont {Chowdhury}, \citenamefont {Hayakawa},\ and\ \citenamefont
  {Izubuchi}}]{Blum:2014oka}%
  \BibitemOpen
  \bibfield  {author} {\bibinfo {author} {\bibfnamefont {T.}~\bibnamefont
  {Blum}}, \bibinfo {author} {\bibfnamefont {S.}~\bibnamefont {Chowdhury}},
  \bibinfo {author} {\bibfnamefont {M.}~\bibnamefont {Hayakawa}}, \ and\
  \bibinfo {author} {\bibfnamefont {T.}~\bibnamefont {Izubuchi}},\ }\href
  {\doibase 10.1103/PhysRevLett.114.012001} {\bibfield  {journal} {\bibinfo
  {journal} {Phys.Rev.Lett.}\ }\textbf {\bibinfo {volume} {114}},\ \bibinfo
  {pages} {012001} (\bibinfo {year} {2015})},\ \Eprint
  {http://arxiv.org/abs/1407.2923} {arXiv:1407.2923 [hep-lat]} \BibitemShut
  {NoStop}%
%%CITATION = ARXIV:1407.2923;%%
\bibitem [{\citenamefont {Pascalutsa}\ and\ \citenamefont
  {Vanderhaeghen}(2010)}]{Pascalutsa:2010sj}%
  \BibitemOpen
  \bibfield  {author} {\bibinfo {author} {\bibfnamefont {V.}~\bibnamefont
  {Pascalutsa}}\ and\ \bibinfo {author} {\bibfnamefont {M.}~\bibnamefont
  {Vanderhaeghen}},\ }\href {\doibase 10.1103/PhysRevLett.105.201603}
  {\bibfield  {journal} {\bibinfo  {journal} {Phys. Rev. Lett.}\ }\textbf
  {\bibinfo {volume} {105}},\ \bibinfo {pages} {201603} (\bibinfo {year}
  {2010})},\ \Eprint {http://arxiv.org/abs/1008.1088} {arXiv:1008.1088
  [hep-ph]} \BibitemShut {NoStop}%
%%CITATION = ARXIV:1008.1088;%%
\bibitem [{\citenamefont {Pascalutsa}\ \emph {et~al.}(2012)\citenamefont
  {Pascalutsa}, \citenamefont {Pauk},\ and\ \citenamefont
  {Vanderhaeghen}}]{Pascalutsa:2012pr}%
  \BibitemOpen
  \bibfield  {author} {\bibinfo {author} {\bibfnamefont {V.}~\bibnamefont
  {Pascalutsa}}, \bibinfo {author} {\bibfnamefont {V.}~\bibnamefont {Pauk}}, \
  and\ \bibinfo {author} {\bibfnamefont {M.}~\bibnamefont {Vanderhaeghen}},\
  }\href {\doibase 10.1103/PhysRevD.85.116001} {\bibfield  {journal} {\bibinfo
  {journal} {Phys.Rev.}\ }\textbf {\bibinfo {volume} {D85}},\ \bibinfo {pages}
  {116001} (\bibinfo {year} {2012})},\ \Eprint {http://arxiv.org/abs/1204.0740}
  {arXiv:1204.0740 [hep-ph]} \BibitemShut {NoStop}%
%%CITATION = ARXIV:1204.0740;%%
\bibitem [{Note1()}]{Note1}%
  \BibitemOpen
  \bibinfo {note} {We use the notation and conventions of~\cite {Peskin:1995ev}
  unless otherwise stated. The metric is mostly minus. The fine-structure
  constant reads $\alpha \equiv {e^2}/({4\pi })\simeq 1/137$. The optical
  theorem for the scattering of scalar particles reads ${\protect \rm
  Im}\protect \tmspace +\thinmuskip {.1667em}{\protect \cal M}(p_1,p_2\to
  p_1,p_2)=2E_{\protect \rm cm}p_{\protect \rm cm}\sigma _{\protect \rm
  tot}(p_1,p_2\to {\protect \rm anything})$, with $E_{\protect \rm cm}$ the
  total center-of-mass energy and $p_{\protect \rm cm}$ the norm of the
  three-momentum of one of the particles in the center-of-mass
  frame.}\BibitemShut {Stop}%
\bibitem [{Note2()}]{Note2}%
  \BibitemOpen
  \bibinfo {note} {We use capital letters to denote `Euclidean' vectors, i.e.\
  the metric in the scalar product of two such vectors is understood to be
  Euclidean.}\BibitemShut {Stop}%
\bibitem [{\citenamefont {Budnev}\ \emph {et~al.}(1971)\citenamefont {Budnev},
  \citenamefont {Chernyak},\ and\ \citenamefont {Ginzburg}}]{Budnev:1971sz}%
  \BibitemOpen
  \bibfield  {author} {\bibinfo {author} {\bibfnamefont {V.~M.}\ \bibnamefont
  {Budnev}}, \bibinfo {author} {\bibfnamefont {V.~L.}\ \bibnamefont
  {Chernyak}}, \ and\ \bibinfo {author} {\bibfnamefont {I.~F.}\ \bibnamefont
  {Ginzburg}},\ }\href {\doibase 10.1016/0550-3213(71)90340-3} {\bibfield
  {journal} {\bibinfo  {journal} {Nucl.Phys.}\ }\textbf {\bibinfo {volume}
  {B34}},\ \bibinfo {pages} {470} (\bibinfo {year} {1971})}\BibitemShut
  {NoStop}%
%%CITATION = NUPHA,B34,470;%%
\bibitem [{Note3()}]{Note3}%
  \BibitemOpen
  \bibinfo {note} {In the notation of~\cite {Pascalutsa:2012pr}, ${\protect
  \cal M}_{\protect \rm TT} = \protect \frac {1}{2}({M}_{++,++}+{M}_{+-,+-} )$
  in terms of the helicity amplitudes. By virtue of the optical theorem, the
  imaginary part of ${\protect \cal M}_{\protect \rm TT}$ is proportional to
  the total unpolarized $\gamma ^*\gamma ^*\to {\protect \rm hadrons}$
  cross-section. For the explicit expression of $R^{\mu \nu }$, see~\cite
  {Budnev:1971sz}.}\BibitemShut {Stop}%
\bibitem [{Note4()}]{Note4}%
  \BibitemOpen
  \bibinfo {note} {One might be able to extend the reach to $|\nu | = \nu _\pi
  $ with methods in the spirit of \cite {Ji:2001wha}.}\BibitemShut {Stop}%
\bibitem [{\citenamefont {Knecht}\ and\ \citenamefont
  {Nyffeler}(2002)}]{Knecht:2001qf}%
  \BibitemOpen
  \bibfield  {author} {\bibinfo {author} {\bibfnamefont {M.}~\bibnamefont
  {Knecht}}\ and\ \bibinfo {author} {\bibfnamefont {A.}~\bibnamefont
  {Nyffeler}},\ }\href {\doibase 10.1103/PhysRevD.65.073034} {\bibfield
  {journal} {\bibinfo  {journal} {Phys.Rev.}\ }\textbf {\bibinfo {volume}
  {D65}},\ \bibinfo {pages} {073034} (\bibinfo {year} {2002})},\ \Eprint
  {http://arxiv.org/abs/hep-ph/0111058} {arXiv:hep-ph/0111058 [hep-ph]}
  \BibitemShut {NoStop}%
%%CITATION = HEP-PH/0111058;%%
\bibitem [{\citenamefont {Francis}\ \emph {et~al.}(2013)\citenamefont
  {Francis}, \citenamefont {Jaeger}, \citenamefont {Meyer},\ and\ \citenamefont
  {Wittig}}]{Francis:2013fzp}%
  \BibitemOpen
  \bibfield  {author} {\bibinfo {author} {\bibfnamefont {A.}~\bibnamefont
  {Francis}}, \bibinfo {author} {\bibfnamefont {B.}~\bibnamefont {Jaeger}},
  \bibinfo {author} {\bibfnamefont {H.~B.}\ \bibnamefont {Meyer}}, \ and\
  \bibinfo {author} {\bibfnamefont {H.}~\bibnamefont {Wittig}},\ }\href
  {\doibase 10.1103/PhysRevD.88.054502} {\bibfield  {journal} {\bibinfo
  {journal} {Phys.Rev.}\ }\textbf {\bibinfo {volume} {D88}},\ \bibinfo {pages}
  {054502} (\bibinfo {year} {2013})},\ \Eprint {http://arxiv.org/abs/1306.2532}
  {arXiv:1306.2532 [hep-lat]} \BibitemShut {NoStop}%
%%CITATION = ARXIV:1306.2532;%%
\bibitem [{\citenamefont {Capitani}\ \emph {et~al.}(2011)\citenamefont
  {Capitani}, \citenamefont {Della~Morte}, \citenamefont {von Hippel},
  \citenamefont {Knippschild},\ and\ \citenamefont {Wittig}}]{Capitani:2011fg}%
  \BibitemOpen
  \bibfield  {author} {\bibinfo {author} {\bibfnamefont {S.}~\bibnamefont
  {Capitani}}, \bibinfo {author} {\bibfnamefont {M.}~\bibnamefont
  {Della~Morte}}, \bibinfo {author} {\bibfnamefont {G.}~\bibnamefont {von
  Hippel}}, \bibinfo {author} {\bibfnamefont {B.}~\bibnamefont {Knippschild}},
  \ and\ \bibinfo {author} {\bibfnamefont {H.}~\bibnamefont {Wittig}},\
  }\href@noop {} {\bibfield  {journal} {\bibinfo  {journal} {PoS}\ }\textbf
  {\bibinfo {volume} {LATTICE2011}},\ \bibinfo {pages} {145} (\bibinfo {year}
  {2011})},\ \Eprint {http://arxiv.org/abs/1110.6365} {arXiv:1110.6365
  [hep-lat]} \BibitemShut {NoStop}%
%%CITATION = ARXIV:1110.6365;%%
\bibitem [{\citenamefont {Fritzsch}\ \emph {et~al.}(2012)\citenamefont
  {Fritzsch}, \citenamefont {Knechtli}, \citenamefont {Leder}, \citenamefont
  {Marinkovic}, \citenamefont {Schaefer} \emph {et~al.}}]{Fritzsch:2012wq}%
  \BibitemOpen
  \bibfield  {author} {\bibinfo {author} {\bibfnamefont {P.}~\bibnamefont
  {Fritzsch}}, \bibinfo {author} {\bibfnamefont {F.}~\bibnamefont {Knechtli}},
  \bibinfo {author} {\bibfnamefont {B.}~\bibnamefont {Leder}}, \bibinfo
  {author} {\bibfnamefont {M.}~\bibnamefont {Marinkovic}}, \bibinfo {author}
  {\bibfnamefont {S.}~\bibnamefont {Schaefer}},  \emph {et~al.},\ }\href
  {\doibase 10.1016/j.nuclphysb.2012.07.026} {\bibfield  {journal} {\bibinfo
  {journal} {Nucl.Phys.}\ }\textbf {\bibinfo {volume} {B865}},\ \bibinfo
  {pages} {397} (\bibinfo {year} {2012})},\ \Eprint
  {http://arxiv.org/abs/1205.5380} {arXiv:1205.5380 [hep-lat]} \BibitemShut
  {NoStop}%
%%CITATION = ARXIV:1205.5380;%%
\bibitem [{\citenamefont {Della~Morte}\ \emph {et~al.}(2005)\citenamefont
  {Della~Morte}, \citenamefont {Hoffmann}, \citenamefont {Knechtli},
  \citenamefont {Sommer},\ and\ \citenamefont {Wolff}}]{DellaMorte:2005rd}%
  \BibitemOpen
  \bibfield  {author} {\bibinfo {author} {\bibfnamefont {M.}~\bibnamefont
  {Della~Morte}}, \bibinfo {author} {\bibfnamefont {R.}~\bibnamefont
  {Hoffmann}}, \bibinfo {author} {\bibfnamefont {F.}~\bibnamefont {Knechtli}},
  \bibinfo {author} {\bibfnamefont {R.}~\bibnamefont {Sommer}}, \ and\ \bibinfo
  {author} {\bibfnamefont {U.}~\bibnamefont {Wolff}},\ }\href {\doibase
  10.1088/1126-6708/2005/07/007} {\bibfield  {journal} {\bibinfo  {journal}
  {JHEP}\ }\textbf {\bibinfo {volume} {0507}},\ \bibinfo {pages} {007}
  (\bibinfo {year} {2005})},\ \Eprint {http://arxiv.org/abs/hep-lat/0505026}
  {arXiv:hep-lat/0505026 [hep-lat]} \BibitemShut {NoStop}%
%%CITATION = HEP-LAT/0505026;%%
\bibitem [{\citenamefont {Dai}\ and\ \citenamefont
  {Pennington}(2014)}]{PhysRevD.90.036004}%
  \BibitemOpen
  \bibfield  {author} {\bibinfo {author} {\bibfnamefont {L.-Y.}\ \bibnamefont
  {Dai}}\ and\ \bibinfo {author} {\bibfnamefont {M.~R.}\ \bibnamefont
  {Pennington}},\ }\href {\doibase 10.1103/PhysRevD.90.036004} {\bibfield
  {journal} {\bibinfo  {journal} {Phys. Rev. D}\ }\textbf {\bibinfo {volume}
  {90}},\ \bibinfo {pages} {036004} (\bibinfo {year} {2014})}\BibitemShut
  {NoStop}%
\bibitem [{\citenamefont {Engel}\ \emph {et~al.}(2015)\citenamefont {Engel},
  \citenamefont {Giusti}, \citenamefont {Lottini},\ and\ \citenamefont
  {Sommer}}]{Engel:2014cka}%
  \BibitemOpen
  \bibfield  {author} {\bibinfo {author} {\bibfnamefont {G.~P.}\ \bibnamefont
  {Engel}}, \bibinfo {author} {\bibfnamefont {L.}~\bibnamefont {Giusti}},
  \bibinfo {author} {\bibfnamefont {S.}~\bibnamefont {Lottini}}, \ and\
  \bibinfo {author} {\bibfnamefont {R.}~\bibnamefont {Sommer}},\ }\href
  {\doibase 10.1103/PhysRevLett.114.112001} {\bibfield  {journal} {\bibinfo
  {journal} {Phys.Rev.Lett.}\ }\textbf {\bibinfo {volume} {114}},\ \bibinfo
  {pages} {112001} (\bibinfo {year} {2015})},\ \Eprint
  {http://arxiv.org/abs/1406.4987} {arXiv:1406.4987 [hep-ph]} \BibitemShut
  {NoStop}%
%%CITATION = ARXIV:1406.4987;%%
\bibitem [{\citenamefont {Bernstein}\ and\ \citenamefont
  {Holstein}(2013)}]{Bernstein:2011bx}%
  \BibitemOpen
  \bibfield  {author} {\bibinfo {author} {\bibfnamefont {A.}~\bibnamefont
  {Bernstein}}\ and\ \bibinfo {author} {\bibfnamefont {B.~R.}\ \bibnamefont
  {Holstein}},\ }\href {\doibase 10.1103/RevModPhys.85.49} {\bibfield
  {journal} {\bibinfo  {journal} {Rev.Mod.Phys.}\ }\textbf {\bibinfo {volume}
  {85}},\ \bibinfo {pages} {49} (\bibinfo {year} {2013})},\ \Eprint
  {http://arxiv.org/abs/1112.4809} {arXiv:1112.4809 [hep-ph]} \BibitemShut
  {NoStop}%
%%CITATION = ARXIV:1112.4809;%%
\bibitem [{\citenamefont {Olive}\ \emph {et~al.}(2014)\citenamefont {Olive}
  \emph {et~al.}}]{Agashe:2014kda}%
  \BibitemOpen
  \bibfield  {author} {\bibinfo {author} {\bibfnamefont {K.~A.}\ \bibnamefont
  {Olive}} \emph {et~al.} (\bibinfo {collaboration} {Particle Data Group}),\
  }\href {\doibase 10.1088/1674-1137/38/9/090001} {\bibfield  {journal}
  {\bibinfo  {journal} {Chin.Phys.}\ }\textbf {\bibinfo {volume} {C38}},\
  \bibinfo {pages} {090001} (\bibinfo {year} {2014})}\BibitemShut {NoStop}%
%%CITATION = CHPHD,C38,090001;%%
\bibitem [{\citenamefont {Eichmann}\ \emph {et~al.}(2015)\citenamefont
  {Eichmann}, \citenamefont {Fischer},\ and\ \citenamefont
  {Heupel}}]{Eichmann:2015nra}%
  \BibitemOpen
  \bibfield  {author} {\bibinfo {author} {\bibfnamefont {G.}~\bibnamefont
  {Eichmann}}, \bibinfo {author} {\bibfnamefont {C.~S.}\ \bibnamefont
  {Fischer}}, \ and\ \bibinfo {author} {\bibfnamefont {W.}~\bibnamefont
  {Heupel}},\ }\href@noop {} {\  (\bibinfo {year} {2015})},\ \Eprint
  {http://arxiv.org/abs/1505.06336} {arXiv:1505.06336 [hep-ph]} \BibitemShut
  {NoStop}%
%%CITATION = ARXIV:1505.06336;%%
\bibitem [{\citenamefont {Edwards}\ and\ \citenamefont
  {Jo{\'o}}(2005)}]{Edwards:2004sx}%
  \BibitemOpen
  \bibfield  {author} {\bibinfo {author} {\bibfnamefont {R.~G.}\ \bibnamefont
  {Edwards}}\ and\ \bibinfo {author} {\bibfnamefont {B.}~\bibnamefont
  {Jo{\'o}}} (\bibinfo {collaboration} {SciDAC, LHPC, UKQCD}),\ }\href
  {\doibase 10.1016/j.nuclphysbps.2004.11.254} {\bibfield  {journal} {\bibinfo
  {journal} {Nucl.Phys.Proc.Suppl.}\ }\textbf {\bibinfo {volume} {140}},\
  \bibinfo {pages} {832} (\bibinfo {year} {2005})},\ \Eprint
  {http://arxiv.org/abs/hep-lat/0409003} {arXiv:hep-lat/0409003 [hep-lat]}
  \BibitemShut {NoStop}%
%%CITATION = HEP-LAT/0409003;%%
\bibitem [{\citenamefont {L{\"u}scher}\ \emph {et~al.}()\citenamefont
  {L{\"u}scher}, \citenamefont {Schaefer} \emph {et~al.}}]{OpenQCD}%
  \BibitemOpen
  \bibfield  {author} {\bibinfo {author} {\bibfnamefont {M.}~\bibnamefont
  {L{\"u}scher}}, \bibinfo {author} {\bibfnamefont {S.}~\bibnamefont
  {Schaefer}},  \emph {et~al.},\ }\href@noop {} {\enquote {\bibinfo {title}
  {{openQCD}},}\ }\bibinfo {howpublished}
  {\url{http://luscher.web.cern.ch/luscher/openQCD/}}\BibitemShut {NoStop}%
\bibitem [{\citenamefont {Peskin}\ and\ \citenamefont
  {Schroeder}(1995)}]{Peskin:1995ev}%
  \BibitemOpen
  \bibfield  {author} {\bibinfo {author} {\bibfnamefont {M.~E.}\ \bibnamefont
  {Peskin}}\ and\ \bibinfo {author} {\bibfnamefont {D.~V.}\ \bibnamefont
  {Schroeder}},\ }\href@noop {} {\emph {\bibinfo {title} {{An Introduction to
  quantum field theory}}}}\ (\bibinfo  {publisher} {Addison-Wesley},\ \bibinfo
  {address} {Reading, USA},\ \bibinfo {year} {1995})\BibitemShut {NoStop}%
%%CITATION = ISBN-9780201503975 ETC.;%%
\bibitem [{\citenamefont {Ji}\ and\ \citenamefont {Jung}(2001)}]{Ji:2001wha}%
  \BibitemOpen
  \bibfield  {author} {\bibinfo {author} {\bibfnamefont {X.}~\bibnamefont
  {Ji}}\ and\ \bibinfo {author} {\bibfnamefont {C.}~\bibnamefont {Jung}},\
  }\href {\doibase 10.1103/PhysRevLett.86.208} {\bibfield  {journal} {\bibinfo
  {journal} {Phys.Rev.Lett.}\ }\textbf {\bibinfo {volume} {86}},\ \bibinfo
  {pages} {208} (\bibinfo {year} {2001})},\ \Eprint
  {http://arxiv.org/abs/hep-lat/0101014} {arXiv:hep-lat/0101014 [hep-lat]}
  \BibitemShut {NoStop}%
%%CITATION = HEP-LAT/0101014;%%
\end{thebibliography}%
%%%%%%%%%%%%%%%%%%%%%%%%%%%%%%%%%%%%%%%%%%%%%%%%%%%%%%%%%%%%%%%%

\end{document}